\documentclass[useAMS,usenatbib]{mn2e}

\usepackage{amssymb}
\usepackage[english]{babel}
\usepackage{url}
\usepackage{amssymb}
\usepackage{graphics}
\usepackage{graphicx}

\usepackage[usenames]{color}

\newcommand{\be}{\begin{equation}}
\newcommand{\ee}{\end{equation}}
\newcommand{\ba}{\begin{eqnarray}}
\newcommand{\ea}{\end{eqnarray}}
\newcommand{\bc}{\begin{center}}
\newcommand{\ec}{\end{center}}

\title[Is there room for high-$\eta$ PWNe?]{Is there room for highly magnetized pulsar wind nebulae \\
among 
those non-detected at TeV?
}
\author[Mart\'in, Torres, Cillis \& de O\~na Wilhelmi]{J. Martin$^{1}$, D. F. Torres$^{1,2}$, A. Cillis$^{3}$ \& E. de O\~na Wilhelmi$^{1}$\\
$^1$Institute of Space Sciences (IEEC-CSIC), Campus UAB,  Torre C5, 2a planta, 08193 Barcelona, Spain \\
$^2$Instituci\'o Catalana de Recerca i Estudis Avan\c{c}ats (ICREA) Barcelona, Spain\\
$^3$Instituto de Astronom\'ia y F\'isica del Espacio,  Casilla de Correo 67 - Suc. 28 (C1428ZAA), Buenos Aires, Argentina \\
}

\begin{document}

\date{}

%\pubyear{2013}

\maketitle

\label{firstpage}

\begin{abstract}

We make a time-dependent characterization of pulsar wind nebulae (PWNe) surrounding some of the highest spin-down pulsars that have not yet been
detected at TeV. Our aim is assessing their possible level of magnetization. We analyze the nebulae driven by J2022+3842 in G76.9+1.0, J0540-6919
in N158A (the Crab twin), J1400--6325 in G310.6--1.6, and J1124--5916 in G292.0+0.18, none of which have been found at TeV energies. For
comparison we refer to published models of G54.1+0.3, the Crab nebula, and develop a model for N157B in the Large Magellanic Cloud (LMC). We
conclude that further observations of N158A could lead to its detection at VHE. According to our model, a FIR energy density of 5 eV cm$^{-3}$
could already lead to a detection in H.E.S.S. (assuming no other IC target field) within 50 hours of exposure and just the CMB inverse Compton
contribution would produce VHE photons at the CTA sensitivity. We also propose models for G76.9+1.0, G310.6--1.6 and G292.0+1.8 which suggest
their TeV detection in a moderate exposure for the latter two with the current generation of Cherenkov telescopes. We analyze the possibility
that these PWNe are highly magnetized, where the low number of particles explains the residual detection in X-rays and their lack of detection at
TeV energies.

\end{abstract}


\section{Introduction}

The spectral energy distribution (SED) of the pulsar wind nebulae (PWNe) of the highest spin-down powered pulsars is diverse. In particular,
luminous pulsars such as Crab ($L_{sd}=4.5 \times 10^{38}$ erg s$^{-1}$) and the Large Magellanic Cloud (LMC) J0537-6910 in N157B
($L_{sd}=4.9 \times 10^{38}$ erg s$^{-1}$) are TeV detected, as are others with spin-down power in the order of several 10$^{37}$ erg s$^{-1}$.
However, several  PWNe with pulsars similarly luminous, are not. Why? Do they have significantly different interstellar environment, injection,
or nebular magnetization?

The X-ray luminosity efficiency of these high-spin down pulsars also presents a large range. A notable case is G76.9+1.0 for which the X-ray
efficiency is $L_{X}/L(t) \thicksim 2.4 \times 10^{-4} D_{10}^2$, where $D_{10}^2$ is the distance in units of 10 kpc \citep{arzou11}.\footnote{The
spin-down of the pulsar in G76.9+1.0 has been recently re-assessed due to a new measurement of the period (see the discussion below). While
it is now lower than earlier claimed, it still qualifies as one the most energetic pulsars we know.} This and similar cases are challenging for
PWNe spectral models since they imply  an inefficient acceleration of high energy electrons in order to fit the X-ray luminosity. For these
cases, \citet{arzou08} suggested that the pulsar wind has a high magnetization factor, arguing that because particle-dominated winds are
necessary for efficient conversion of wind to synchrotron power, PWNe with high magnetization would lead to dim X-ray PWNe. Thus, high-$\eta$
(high magnetic fraction) models point to an interesting alternative for the interpretation of PWNe, which, despite their high spin-down, lack TeV
emission and have weak X-ray counterparts. These PWNe would be different to TeV detected ones. Except CTA 1, for which the magnetization reaches
almost to equipartition,  all TeV-PWNe with characteristic ages of 10 kyr or less can be described with an spectral model with low $\eta$, and are thus
strongly particle dominated  \citep{torr14}. 

An interesting case is that of G292.0+0.18, for which the central powering pulsar, J1124--5916, has essentially the same $P$, $\dot P$ (up to
three significant decimal places) than J1930+1852, which powers G54.1+0.3. The distance for both nebulae is also similar ($\sim$6 kpc). Whereas
the latter is a TeV source, and modeled as particle dominated PWN (e.g., \citealt{tana11}, \citealt{torr14}), the former is not (at least at the
level in which it has been covered in the Galactic Plane observations by H.E.S.S. \citep{carri13}. With the same spin-down power and located at a
similar Galactic distance, it seems that the flux at TeV energies depends on other factors such the environment (the FIR density, for instance)
or the nebula magnetization. Is then G292.0+0.18 simply like G54.1+0.3 but subject to a stronger magnetization?

\citet{tana13} have also analyzed several PWNe which have been undetected at TeV.\footnote{For differences between their model and ours see the
discussion in \citet{mart12} and \citet{torr14}. Their magnetic field evolution does not consider losses in magnetic energy due to expansion, and
thus their magnetization values are lower than ours typically by a factor 2--3.} However, they assumed a fixed low magnetization
($3 \times 10^{-3}$) compatible with usual particle dominated nebulae that have been detected at TeV to describe them. In this work, we explore
the phase space of PWNe models also in magnetization, in order to distinguish whether there is preference for the existence of highly magnetized
nebulae (or at least, for nebula with magnetization close to equipartition) among those not yet seen at TeV.

\section{Spectral model}

The code we use solves the time-dependent diffusion-loss equation for the electrons in the PWN, and includes the energy losses due to
synchrotron, IC, adiabatic and Bremsstrahlung processes, and escape due to Bohm diffusion (see \citealt{mart12,torr13,torr13b} for details). We
shall refer with $Q(\gamma,t)$ to the injection function, generally assumed as a broken power law with $\gamma_b$ being the energy of the break
(in Lorentz factor units) and $\alpha_l$ and $\alpha_h$, the low and high energy indices respectively. The normalization of the injection is
computed using the spin-down luminosity of the pulsar $L(t)$ and the magnetic fraction $\eta$
\begin{equation}
(1-\eta)L(t)=\int_{\gamma_{min}}^{\gamma_{max}} \gamma m_e c^2 Q(\gamma,t) \mathrm{d}\gamma.
\end{equation}
The evolution of the magnetic field is described by (e.g., \citealt{paci73})
\begin{equation}
\label{mag}
B(t)=\frac{1}{R_{PWN}^2}\left[6 \eta \int_0^t L(t')R_{PWN}(t')\mathrm{d}t' \right]^{1/2}.
\end{equation}
The spin-down power evolution is deduced assuming that the pulsar has the same behavior as a spinning magnetic dipole with braking index $n$.  
\begin{figure}
\begin{center}
\includegraphics[scale=0.5]{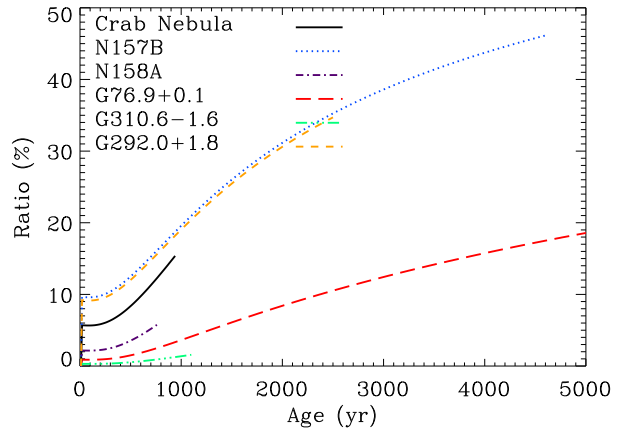}
\caption{Ratio of the PWN radii resulting from the analytical and numerical models as commented in the text, for some of the nebulae studied.}
\label{radcomp}
\end{center}
\end{figure}

We refer with $\tau_0=2\tau_c/(n-1)-t_{age}$ to the initial spin-down age, and with $\tau_c$ to the characteristic age of the pulsar. The maximum
energy of the particles is calculated by the requirement that their Larmor radius is smaller than the termination shock radius. The radius of the
nebula is computed assuming free expansion of the shell in the interior of the supernova remnant (SNR). If we consider that the spin-down
luminosity of the pulsar is constant, it evolves as \citep{swa01}
$
R_{PWN}(t)=0.839 \left({L_0 t}/{E_0} \right)^{1/5}V_0 t,
$
with
$
V_0=\sqrt{{10E_0}/{3M_{ej}}}. 
$
We shall refer to this radius as $R_{analytical}$.
This value is obtained by assuming that all the mechanical energy of the explosion with energy $E_0$ is transformed into kinetic energy. $M_{ej}$
is the ejected mass in the explosion. {$L_0$ is the initial spin-down luminosity of the PSR.} In some of the PWNe that we study in this work, the
spin-down power of the pulsar is very high and its variation in time cannot be neglected. For this reason, we solve numerically the energy
balance equation
\begin{equation}
\frac{d}{dt}(4\pi R_{PWN}^3 P)=L(t)-4\pi R_{PWN}^2(t) P(t) \left(\frac{dR_{PWN}}{dt}\right)
\end{equation}
assuming that the pressure $P(t)$ of the PWN at the contact discontinuity with the expanding ejecta is given by (equation A7 in \citealt{swa01})
\begin{equation}
P(t)=\frac{3}{25}\rho_{ej}(t)\left(\frac{R_{PWN}(t)}{t}\right)^2,
\end{equation}
which depends on the density of the SNR ejecta $\rho_{ej}$. Fig. \ref{radcomp} shows the fractional deviation between the radii, i.e.,
$(R_{analytical}-R_{numerical})/R_{analytical} \times 100$, using the parameters obtained  of Table 1. We observe that the difference in radius
goes from $\sim$2\% (for G310.6-1.6) to $\sim$46\% (for N157B), being larger for when the spin-down luminosity is constant. Taking into account
that the magnetic field depends on the radius as $B \sim R_{PWN}^{-3/2}$, this changes the magnetic field of the nebula, and consequently, its
synchrotron flux. Another affected parameter in the fits is the ejected mass of the progenitor star, because the velocity of the ejecta behaves
with the mass as $V_0 \sim M_{ej}^{-1/2}$. This means that with the numerical solution, the ejected mass needed to reach a given radius $R_{PWN}$
has to be smaller, due to the decreasing spin-down power in time. Regarding the SSC radiation, we assume that the synchrotron ball generating the
multi-frequency radiation from each of the PWN has the same radius at all energies, and is equal to the radius of the PWN itself.

In order to fit the spectra, we fix from observations as many parameters as possible. Generally, we fix the period ($P$), period derivative
($\dot{P}$), braking index ($n$), age of the system ($t_{age}$), initial spin-down age ($\tau_0$), spin-down luminosity ($L(t)$), distance ($d$),
radius of the PWN ($R_{PWN}$), minimum energy at injection ($\gamma_{min}$), and FIR and NIR temperatures ($T_{FIR}$, $T_{NIR}$). The rest of
parameters are fitted or derived from the others.

\section{Large Magellanic Cloud's N157B \& N158A}

\begin{figure*}
\begin{center}
\includegraphics[scale=0.46]{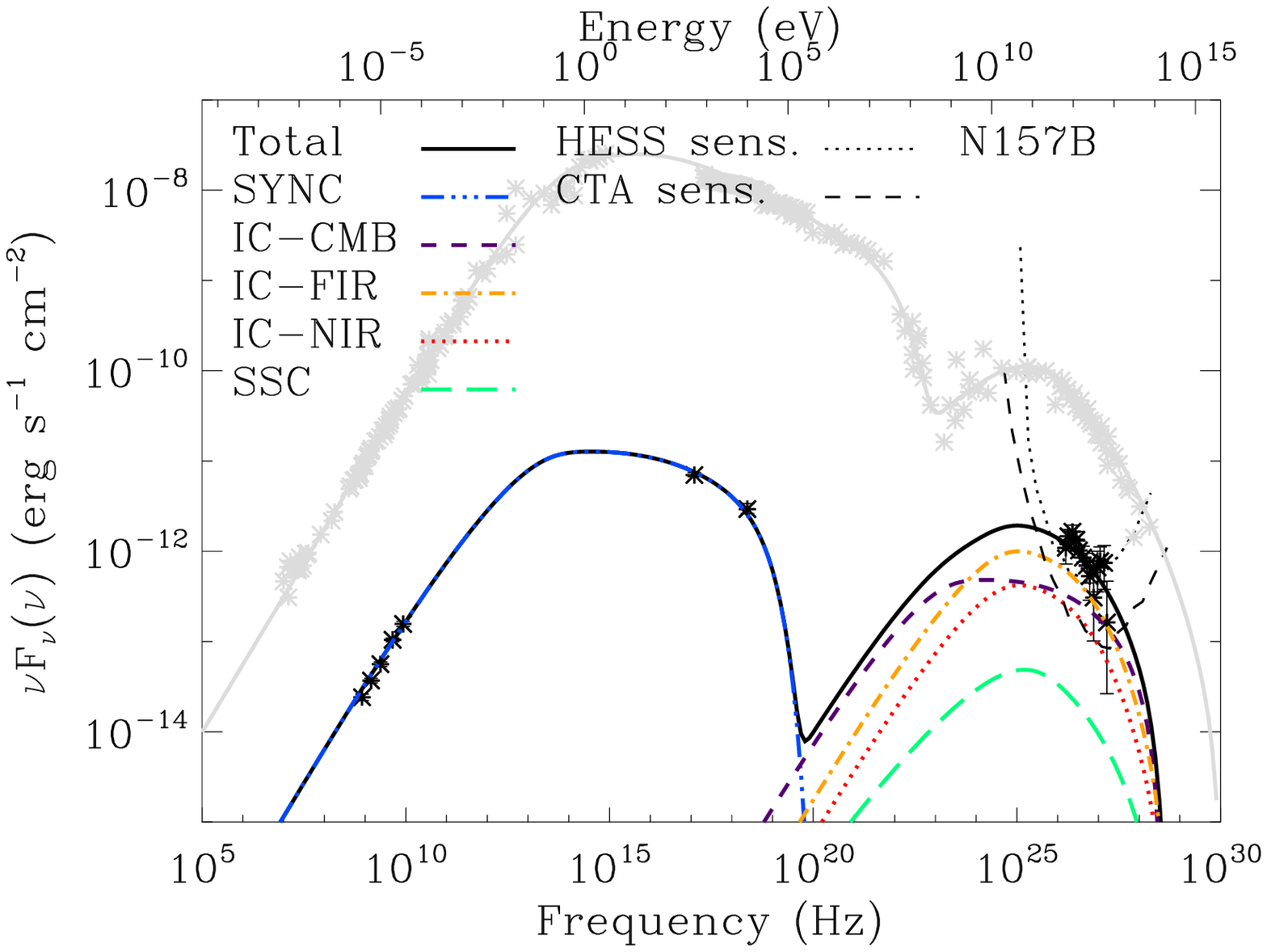}
\hspace{0.8cm}
\includegraphics[scale=0.46]{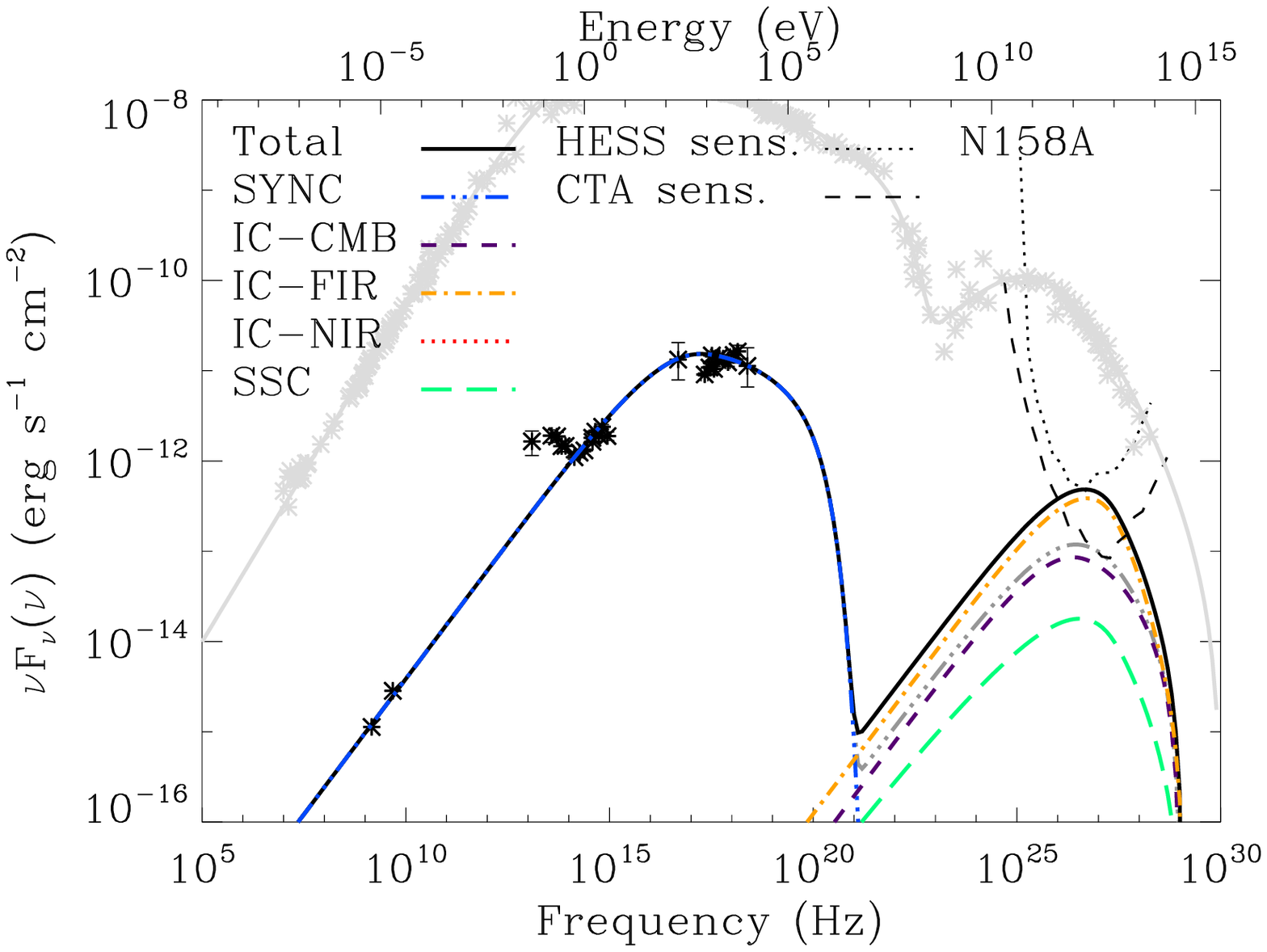}
%\hspace{0.8cm}
%\includegraphics[scale=0.46]{n158a_cta.ps}
%\includegraphics[scale=0.46]{g310_model1.ps}
%\hspace{0.8cm}
%\includegraphics[scale=0.46]{g310_model2.ps}
\caption{Nebulae in the LMC. Left panel: Spectral fit for the N157B PWN. The fluxes and the fit of the Crab Nebula is overplotted in
grey for comparison. We plot also the sensitivity curves of H.E.S.S. and CTA for an exposure time of 50 hours. The data points are
obtained from: \protect\cite{lazen00} (radio), \protect\cite{chen06} (X-rays), \protect\cite{abra12} (VHE). Right panel: Spectral fit for
the N158A PWN to reach H.E.S.S. (in solid black) and CTA (in triple-dot dashed grey) sensitivities. The data points are obtained from:
\protect\cite{man93b} (radio), \protect\cite{migna12} (infrared \& optical), \protect\cite{kaa01}, \protect\cite{camp08} (X-rays).
%Bottom panels: Spectral fits for G310.6--1.6 PWN. Data points
%are taken from \protect\cite{ren10}.
}
\label{lmcpwn}
\end{center}
\end{figure*}

N157B is located in the LMC and it was the first extragalactic PWN detected in gamma rays \citep{abra12}. Its pulsar, PSR
J0537-6910, has a spin-down power of 4.9 $\times$ 10$^{38}$ erg s$^{-1}$ \citep{man05}. \cite{lazen00} did radio observations of
this PWN  using the Australia Telescope Compact Array ({\it ATCA}), obtaining a spectral index of --0.19. \citet{mice09} did
infrared observations using the {\it Spitzer} telescope but reported no infrared counterpart (no bright SNR). Studying the gas and dust
properties of the vicinity, they deduced that the mass of the progenitor star should not be higher than 25M$_{\odot}$. In X-rays,
N157B was observed with {\it ASCA} and {\it ROSAT} \citep{wang98}, and  \citet{wang01} detected the PWN with {\it Chandra}.
\citet{chen06} analyzed the spectrum of N157B and the pulsar PSR J0537-6910 in X-rays. The spectrum of the entire remnant is fitted
with a dominant non-thermal component (a power-law with a spectral index of 2.29 and an unabsorbed flux of 1.4 $\times$ 10$^{-11}$
erg s$^{-1}$ cm$^{-2}$) and a thermal component (a NEI model with a temperature of 0.72 keV and an unabsorbed flux of 7 $\times$
10$^{-12}$ erg s$^{-1}$ cm$^{-2}$).

In our study, we use the estimated distance of 48 kpc, see \cite{abra12}. There are two gas bubbles in the vicinity of
N157B which contribute to the far-infrared (FIR) photon fields: 30 Doradus complex and the OB association
LH99. From the infrared observations done by \cite{inde09}, \cite{abra12} modelled the infrared emission as a black body with
energy density of 8.9 eV cm$^{-3}$ and a temperature of 88 K for the LH99 association, and 2.7 eV cm$^{-3}$ and 80 K for 30 Doradus.
They consider these values as an upper limit since the unprojected distance between these objects is unknown.

Figure \ref{lmcpwn} (left panel) shows the fit we obtain for N157B. We assume the radius for the PWN given by \citet{lazen00}, i.e., 10.6
pc for a distance of 48 kpc. The PWN shell is not very well defined and some small contribution of the SNR could be included. In
this first model, we assumed an age of 4600 yr, which is consistent with the Sedov age of the SNR given by \citet{wang98} ($\sim$5
kyr) and an ejected mass of 20M$_\odot$, corresponding to the lower limit in the ejected
mass given by \citet{chen06}. The electron injection has a low (high) energy index of 1.5 (2.75) and the energy break is located at
$\gamma=10^6$ ($\sim$511 GeV). From the synchrotron part of the spectrum, we inferred a magnetic field of 13 $\mu$G and a magnetic
fraction of 0.006. The energy density of the target photon fields, enhanced due to the near presence of LH99 and 30 Doradus, results
in our fits much below the upper limits given by \citet{abra12}, i.e., 0.7 and 0.3 eV cm$^{-3}$, respectively. 

If instead we assume the energy densities given by \citet{abra12}, we need to consider a lower age of 2.5 kyr to fit the TeV data.
Considering the lower limit on the ejected mass given by \citet{chen06}, then the radius decreases until 3.7 pc. Regarding the synchrotron
spectrum, the magnetic field reaches 35$\mu$G and $\eta$=0.01. The intrinsic energy break changes to $\gamma_b=2 \times 10^5$
($\sim$102 GeV) and the injection slopes change slightly ($\alpha_l$=1.5, $\alpha_h$=2.6).
%
%The derived Sedov age of the N157B remnant is 5 kyr \citep{wang98}, which is consistent
%with the characteristic age of the N157B pulsar (see Table 1). In our model, in order to compute the possible age of N157B, we can use
%the radius measured by \citet{lazen00} of 10.6 pc. Assuming the lower limit in the ejected mass given by \citet{chen06}, the
%resulting age (4600 yr) is consistent with the remnant age derived by \citet{wang98}. Model 1 for N157B implements these
%restrictions. In model 2 we fix instead the ejected mass at 20M$_\odot$ and the FIR upper limit photon fields to those computed by
%\citet{abra12}. In order to fit the IC spectrum, we need to assume an age of 2500 yr which implies a radius of 3.7 pc. 
%
The value obtained for the radius in the latter model is only $\sim$50\% higher than the radius observed in X-rays. This difference
is small in comparison with other cases. For example, for the Crab nebula, we see that the radius in the radio band is $\sim$2 pc and in
X-rays $\sim$0.6 pc. As the shell is not well defined, the radius measured by \citet{lazen00} could include parts of the remnant, but the
relation between the PWN radius in X-rays and the radius in the radio band seems to be more similar to the Crab nebula case. \citet{swa04}
suggested that N157B PWN could be interacting with the reverse shock of the SNR in a very initial phase, explaining its elongated
morphology. In any case, we find that N157B is a luminous particle dominated nebula. \\

%%%%%%%%%%%%%%%%%%%%%
%\subsection{N158A}
%%%%%%%%%%%%%%%%%%%%%

N158A, known as the Crab twin, is also located in the LMC but has not been detected at TeV yet. This PWN is powered by the pulsar PSR B0540--69,
which has been observed in radio, infrared, optical and X-ray bands. The period of this pulsar is 50.5 ms \citep{sew84} and the period
derivative is 4.7 $\times$ 10$^{-13}$ s s$^{-1}$ \citep{livi05}. The resulting spin-down luminosity is then 1.5 $\times$ 10$^{38}$ erg
s$^{-1}$. The diameter of N158A is 1.4 pc, as obtained from radio observations  \citep{man93b}. The distance to PSR B0540--69 has been
estimated as $\sim$49 kpc \citep{sew84,tay93b,slo07}. 
An age of 760 yr is deduced through measurements of the expansion velocity of the SNR shell in the optical spectral range
\citep{rey85,kir89}. There is no observational measurement of the ejected mass in N158A, and we have left this parameter free in our model.
The resulting ejected mass in our fits is 25M$_\odot$. According to \citet{heger03}, this mass is at the limit for neutron star creation,
which can grow with the quantity of helium in the core of the star and the energy of the supernova explosion. 
In the infrared, \citet{cara92} did a high-resolution observation of N158A using the European Southern Observatory
New Technology Telescope ({\it ESO-NTT}) and concluded that the progenitor of the SNR could have belonged to the same generation of
young stars in 30 Doradus \citep{cara92,kir89}. \citet{will08} did not find evidence of infrared emission from the SNR, but they
inferred a mass of 20--25M$_{\odot}$ for the progenitor star. PSR B0540--69 is one of the few pulsars with optical pulsations and
polarized emission. Its optical spectrum is well fitted by a power-law, but joining it with the X-ray spectrum, a double break is
required \citep{migna12}. The braking index for PSR B0540--69 is 2.08 \citep{kaa01}. A high-resolution X-ray observation was done with
{\it Chandra} \citep{gott00,kaa01} and there is also a compilation of the observations done with {\it RXTE}, {\it Swift} and
{\it INTEGRAL} in the work by \citet{camp08}. The flux obtained for the PWN is $\sim$8 $\times$ 10$^{-11}$ erg s$^{-1}$ cm$^{-2}$. There
is no detection of the PWN at VHE.

For N158A, the injection spectrum resulting from our fit is a broken power-law with break at a large energy $\gamma=3 \times 10^7$
($\sim$15.3 TeV) and a low (high) energy spectral index of 1.8 (2.6). The synchrotron component is fitted with a magnetic field of 32
$\mu$G. The magnetic fraction in this case is  low  ($\eta=0.0007$). 
Due to the lack of information
on the FIR and near-infrared/optical (NIR) fields, we assume a FIR field with a temperature of 80 K and compute the energy
density needed for the PWN to be detected by H.E.S.S. or CTA. For H.E.S.S., a minimum energy density of 5 eV cm$^{-3}$ is
required to be detected in a 50 hours observation, according to the sensitivity curve used here. 
For CTA, an energy density of 0.2 eV cm$^{-3}$ would be
enough to allow detection, which foresees its identification in case our model is correct. Both models are shown in Fig.
\ref{lmcpwn} (right panel) and their parameters are given in Table 1.

The NIR photon field could also be important depending on the density of nearby stars in the N158A field
and could enhance the TeV yield, at the same time reducing the required FIR densities for detection. 

We have also investigated highly magnetized models in which the detection of N158A is not possible even with CTA, 
unless the energy density of the FIR increases up to $\sim$500 eV cm$^{-3}$ (assuming that there is no NIR contribution). The injection
function in such models has an energy break of $\gamma_b=6 \times 10^7$, and a low (high) spectral index is 1.45 (2.4). Taking into account that
the maximum energy at injection is $\gamma_{max}=1.2 \times 10^8$, a simple power-law model with an index of 1.45 could also be
compatible with this fit. Here, we obtain a highly magnetized nebula with a magnetic fraction of 0.9 and an extreme magnetic
field of 1.15 mG. But whereas the radio and the infrared data are fitted similarly well to particle dominated models, the predicted X-ray flux
of these models is not quite in agreement with data. This fact and the extreme values of the parameters we have just quoted make a high
$\eta$ model unlikely. In equipartition (i.e., $\eta=0.5$), the radio and X-ray flux surpasses the data flux in a factor $\sim$4. In this latter
case, the magnetic field is lower ($B=858$ $\mu$G), but the number of particles is still high to fit the flux.
%Note that with such a high value of $\eta$, we need an extremely low value of the parameter $\epsilon$, the containment factor (the ratio
%between the shock and the nebula radius). This happens because to  ensure particle confinement, we impose that the Larmor radius of the
%particles has to be smaller than the termination shock radius, what implies $\gamma_{max} \propto \epsilon \sqrt{\eta L(t)}$.

We conclude that N158A  is a particle dominated
nebulae that has been undetected because of sensitivity limitations. 

\section{Powerful Galactic pulsars having nebulae non-detected at TeV yet} 

\subsection{G76.9+1.0}

G76.9+1.0 hosts the pulsar PSR J2022+3842. The period and the period derivative of this pulsar was firstly determined by
\citet{arzou11}. They obtained a period of 24 ms and a period derivative of 4.3 $\times 10^{-14}$ s s$^{-1}$, which implies a
spin-down luminosity of 1.2 $\times 10^{38}$ erg s$^{-1}$. This made PSR J2022+3842 the third pulsar with the highest spin-down
known. In later observations with {\it XMM-Newton}, \citet{arumu13}  discovered a factor
2 error in the determination of the pulsar period and period derivative. The new period is then 48 ms and the spin-down luminosity
reduces to 2.96 $\times 10^{37}$ erg s$^{-1}$. 

The remnant was observed in radio using the Very Large Array telescope ({\it VLA}) \citep{lande93}. These authors assume a distance
of 7 kpc, which implies a size of 18$\times$24 pc. The structure of the SNR is dominated by two lobes oriented in the north-south
direction separated by 3 arcmin. The spectral index is 0.62$\pm$0.04. They looked for an infrared counterpart using IRAS data but
none was found. \citet{arzou11} observed PSR J2022+3842 in X-rays using {\it Chandra}, obtaining an absorbed
X-ray flux (2--10 keV) of 5.3 $\times$ 10$^{-13}$ erg s$^{-1}$ cm$^{-2}$ and  detecting a very weak PWN with an absorbed flux of
4 $\times$ 10$^{-14}$ erg s$^{-1}$ cm$^{-2}$. 
In this case, there is no TeV detection either, 
and we only have information about the spectrum in X-rays and upper limits in radio using the flux observed for the SNR
radio shell. 

We adopted an age of 5 kyr, which
implies a reasonable ejected mass of 20M$_\odot$, also proposed by \citet{arzou11}. There are no estimations of the age of
the remnant and of the ejected mass. \citet{arzou11} has established an upper limit on the true age of the pulsar depending on the braking index
of $\sim$40 kyr, which is unconstraining.

%We fit three different models, see Fig.  \ref{g76}.  
We use the data simulated by GALPROP \citep{por06} for the energy
densities and temperatures for the FIR and NIR photon fields, essentially, 
diluted black bodies with a temperature of 25 K and an energy density of 0.13 eV cm$^{-3}$ for the FIR field, and a temperature of
3200 K and an energy density of 0.33 eV cm$^{-3}$ for the NIR field. 
As the PWN in
X-rays is very diluted, its shell cannot be distinguished. For this reason, to simulate the expansion of the nebula, we assumed a
ballistic expansion of the SNR radio shell ($R_{SNR}=V_0 t$) and compute the necessary ejected mass. In this case, we obtain a
value of $\sim$20M$_\odot$, which implies a radius of $\sim$6.3 pc. We assume a braking index of 3, which implies a reasonable value for the
initial period for PSR J2022+3842 of 32 ms.

In model 1, see Table 1, we assume a broken power-law injection with a low-energy (high-energy) spectral
index of 1.5 (2.7). The resulting energy break is at $\gamma_b=10^3$. The magnetic field (3.5 $\mu$G) is close to
the average ISM value. The magnetic fraction is 0.0017. The low value of the injection energy break in this model argues for 
a possible simple power-law injection. This is assumed in model 2. In this case, the spectral index is 2.6 and the magnetic
field is 16.6 $\mu$G, with a magnetic fraction of 0.038. Finally, model 3 explores whether G76.9+1.0 could be a highly magnetic
PWN, as speculated previously by \citet{arzou08,arzou11}. The lack of significant observational constraints allows to entertain this
possibility, and we show an example with a magnetic field of
85.2 $\mu$G and a magnetic fraction of $\eta=0.998$.
The injection function in this case is a simple power-law with an spectral
index of 2.65. In order to respect the upper limits in radio, we need to impose a minimum energy at injection for particles of
$\gamma=10^4$ ($\sim$5.1 GeV). The IC contribution decreases with respect the other models, as expected due to the lower contribution of
spin-down energy to particles and the larger synchrotron field, which maximizes their losses. 

The lack of observational data to put sufficient constrains to differenciate the models proposed. In any case, its detection at TeV energies
seems unexpected.

\begin{figure}
\vspace{0.2cm}
\begin{center}
\includegraphics[scale=0.46]{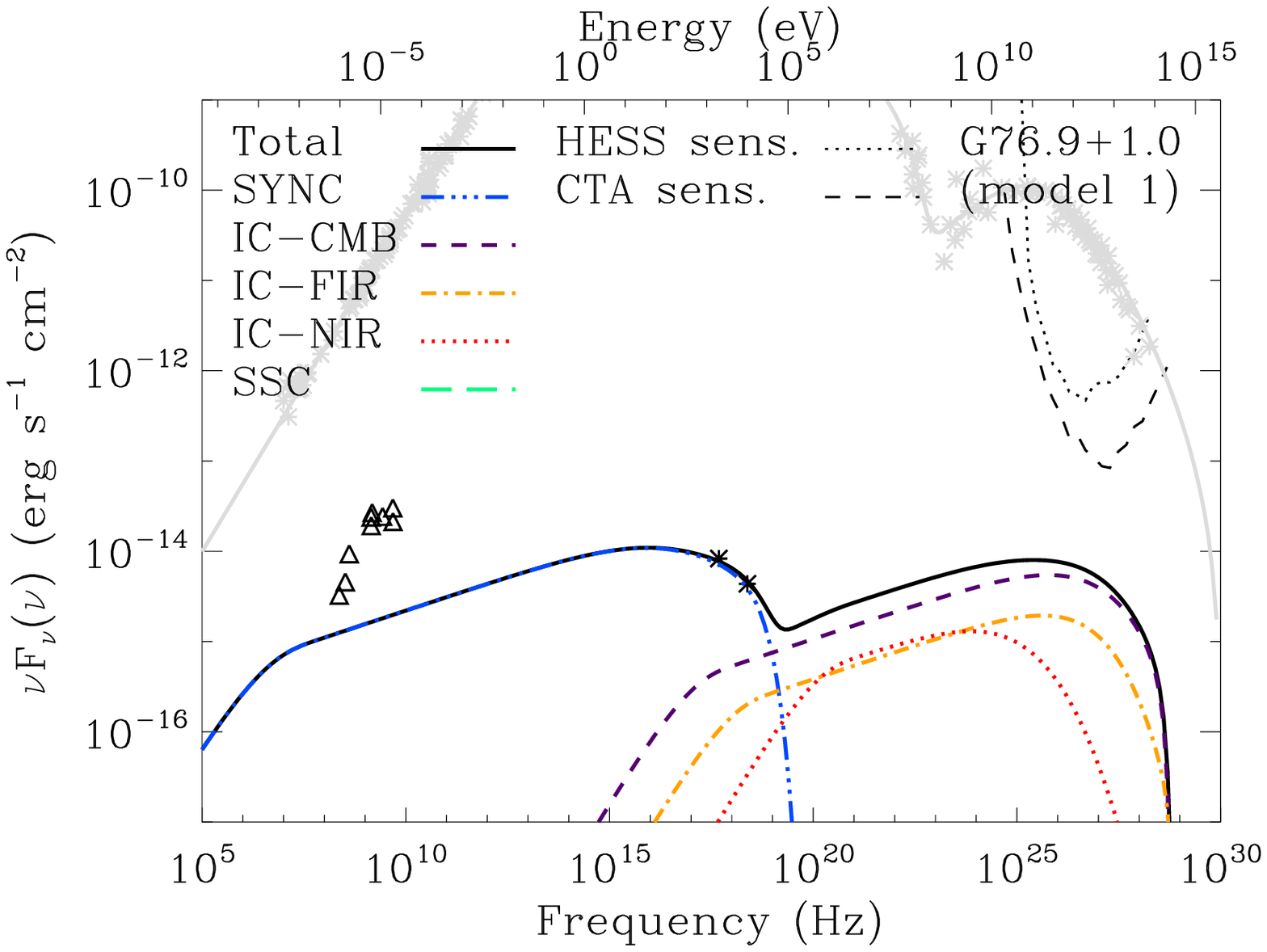}
%\hspace{0.8cm}
\includegraphics[scale=0.46]{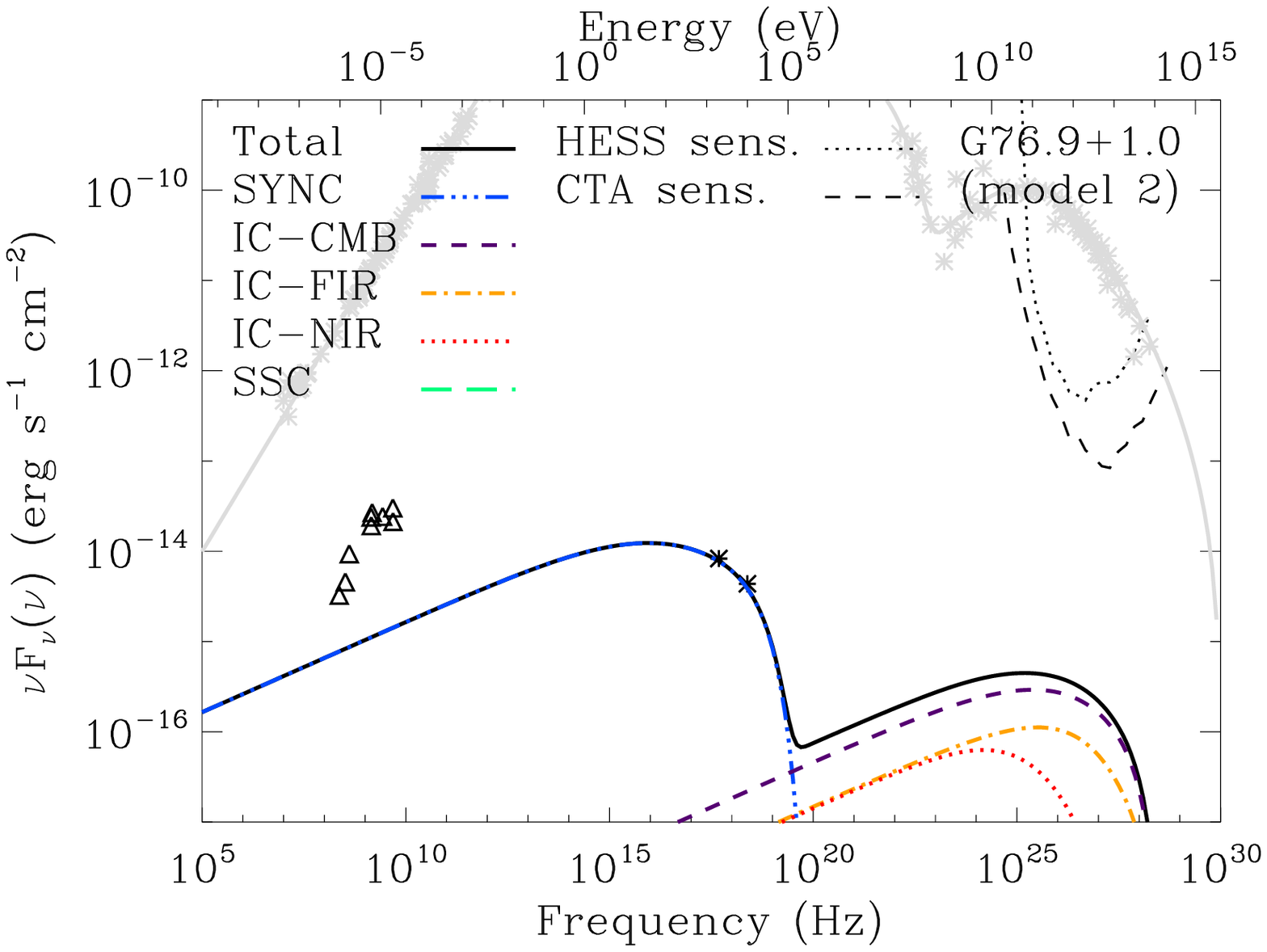}
\includegraphics[scale=0.46]{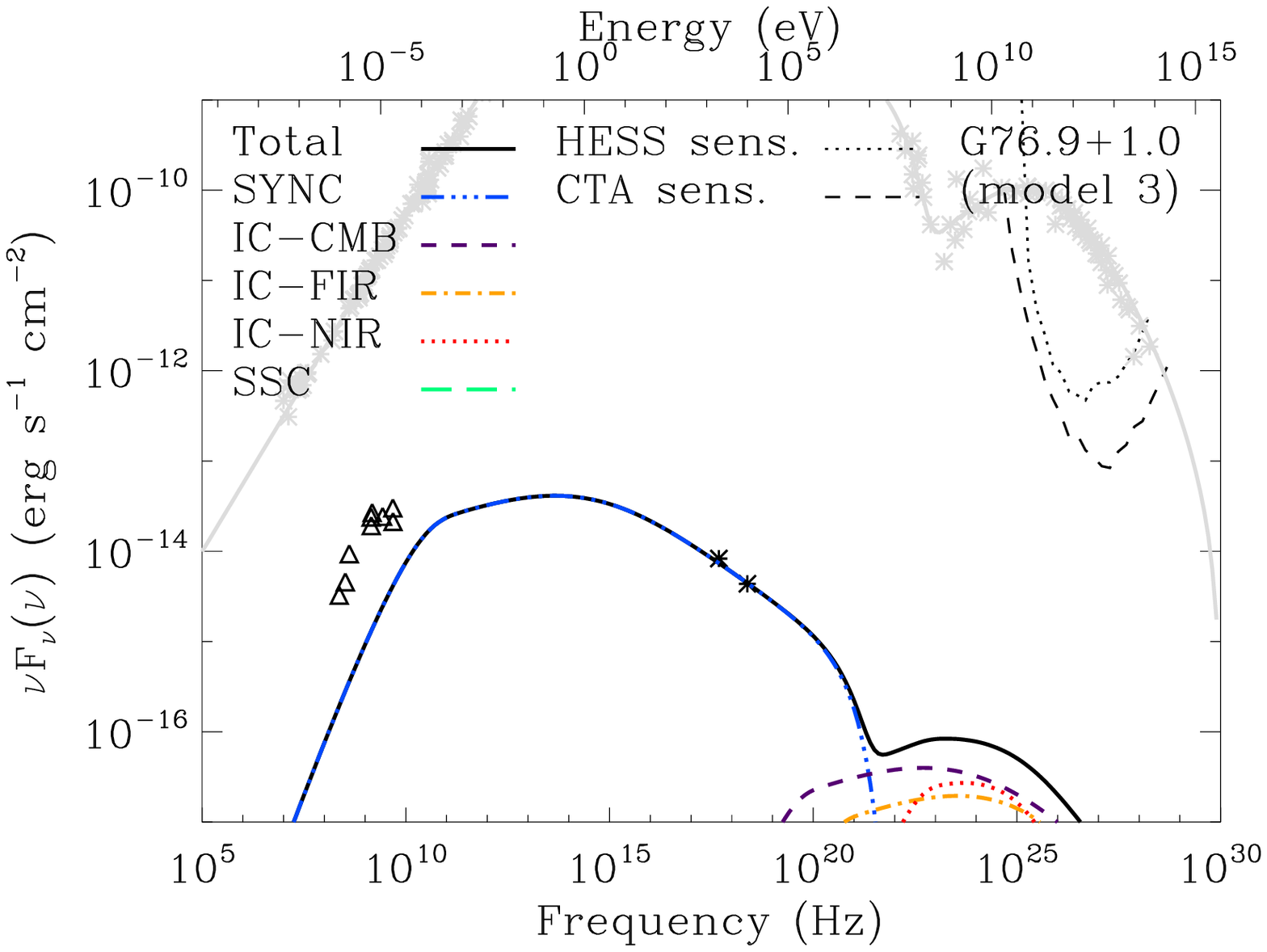}
\caption{Spectral fits for G76.9+1.0 PWN (models 1 to 3, top to bottom). The triangle data points correspond to the radio flux of the radio shell given in
\protect\cite{lande93}, here used as upper limits. The X-ray data is obtained from \protect\cite{arzou11}.}
\label{g76}
\end{center}
\end{figure}

\subsection{G310.6--1.6}

G310.6--1.6 (IGR J14003-6326) was discovered as a soft $\gamma$-ray source in a deep mosaic of the Circinus region done by INTEGRAL
 \citep{keek06}. It was also observed in the {\it Swift} survey of  INTEGRAL sources,
but without conclusions about its origin \citep{mal07}. 
With {\it Chandra} observations, \citet{tom09} fitted the spectrum (0.3 and 10 keV) of the source 
with a power-law with a photon index of $\Gamma=1.82 \pm 0.13$. \citet{ren10} discovered
31.18 ms pulsations using {\it RXTE}, as well as reported the radio detection of PSR J1400-6325 and its nebula. From the {\it RXTE} timing analysis, they
obtained a period derivative for PSR J1400-6325 of 3.89 $\times 10^{-14}$ s s$^{-1}$, which implies an spin-down luminosity of 5.1
$\times 10^{37}$ erg s$^{-1}$ and a characteristic age of 12.7 kyr. There are several estimations of the PWN distance, covering a range
between 6 and 10 kpc. We adopt the value of 7 kpc given in \citet{ren10}.

\citet{ren10} have studied the spectrum of G310.6--1.6, PSR J1400-6325 and its PWN from 0.8 to 100 keV. The spectrum is highly
dominated by the PWN and it is fitted with a broken power-law. The energy break is located at 6 keV and it is probably produced by the
synchrotron cooling of the particles. The spectral index for energies lower (higher) than the energy break is 1.90$\pm$0.10 (2.59$\pm$0.11). The total flux for the PWN at 20--100 keV is 5.3 $\times 10^{-12}$ erg cm$^{-2}$ s$^{-1}$. The PWN flux in radio frequencies
has also been measured, using data from the Molonglo Galactic Plane Survey \citep{mur07} at 843 MHz, as 217.4$\pm$9.4 mJy, as well as from
the Parkes-MIT-NRAO (PMN) survey \citep{gri93,con93} at 4.85 GHz, as 113$\pm$10 mJy. An upper limit of 0.6 mJy
at 2.4 GHz is also established by the Parkes telescope \citep{dun95}.
%The morphology of the nebula is quite surprising since the X-ray PWN is 0.65', while the radio nebula is 44''.8$\times$31''.2. In other
%Crab-like nebulae, we observe that the size of the nebula decreases with energy. 
At TeV energies, G310.6--1.6 was observed by H.E.S.S.
\citep{cha08}, but only an upper limit of 4\% of the Crab Nebula was established \citep{khe08}. 

The spectrum of G310.6-1.6 PWN has been previously studied by \citet{tana13}. They assumed a magnetic fraction of 0.003, an age of
600 yr and a distance of 7 kpc. For these parameters, they obtained an injection with a low (high) energy spectral index of 1.4 (3.0)
with an energy break of $\gamma_b=3 \times 10^6$ and a magnetic field of 17 $\mu$G. They assumed a 0.3 eV cm$^{-3}$ energy density for
the FIR and NIR target photon fields.

In our case, we firstly propose a low magnetized model (model 1), where we assume
that the age of the PWN is 1.1 kyr, which is consistent with the upper limit of 1.9 kyr established by \citet{ren10}, but older than
the one considered in \citet{tana13}. This assumption has been done also taking into account the size of the nebula and a reasonable
ejected mass of 9 M$_\odot$ with a SN energy of 10$^{51}$ erg. \citet{ren10} proposed a subenergetic SN of $5 \times 10^{48}$ erg setting
an ISM density of 0.01 cm$^{-3}$. This implies an ejected mass of 3M$_\odot$ to explain the size of the nebula. This mass is very low for
the ejecta of a star that explodes as a SN. We also prefer to consider the canonical value for the SN explosion energy.

The target photon fields are obtained from those computed by
GALPROP. The fitted black bodies of these photon fields have a temperature of 25 K and 3300 K and an energy density of 0.63 eV cm$^{-3}$
and 1.62 eV cm$^{-3}$ for FIR and NIR, respectively. The obtained magnetic
field is 8.2 $\mu$G and $\eta$=0.0007. The latter is 
the same value we find for the particle dominated models of N158A.
The value of the magnetic field agrees with the lower limit of 6 $\mu$G given by \citet{ren10}.
The intrinsic energy break of the injection in this model is located at $\gamma=2 \times 10^6$ ($\sim$1 TeV). The injection index at low
(high) energies is 1.5 (2.5). 

\begin{figure}
\vspace{0.2cm}
\begin{center}
\includegraphics[scale=0.46]{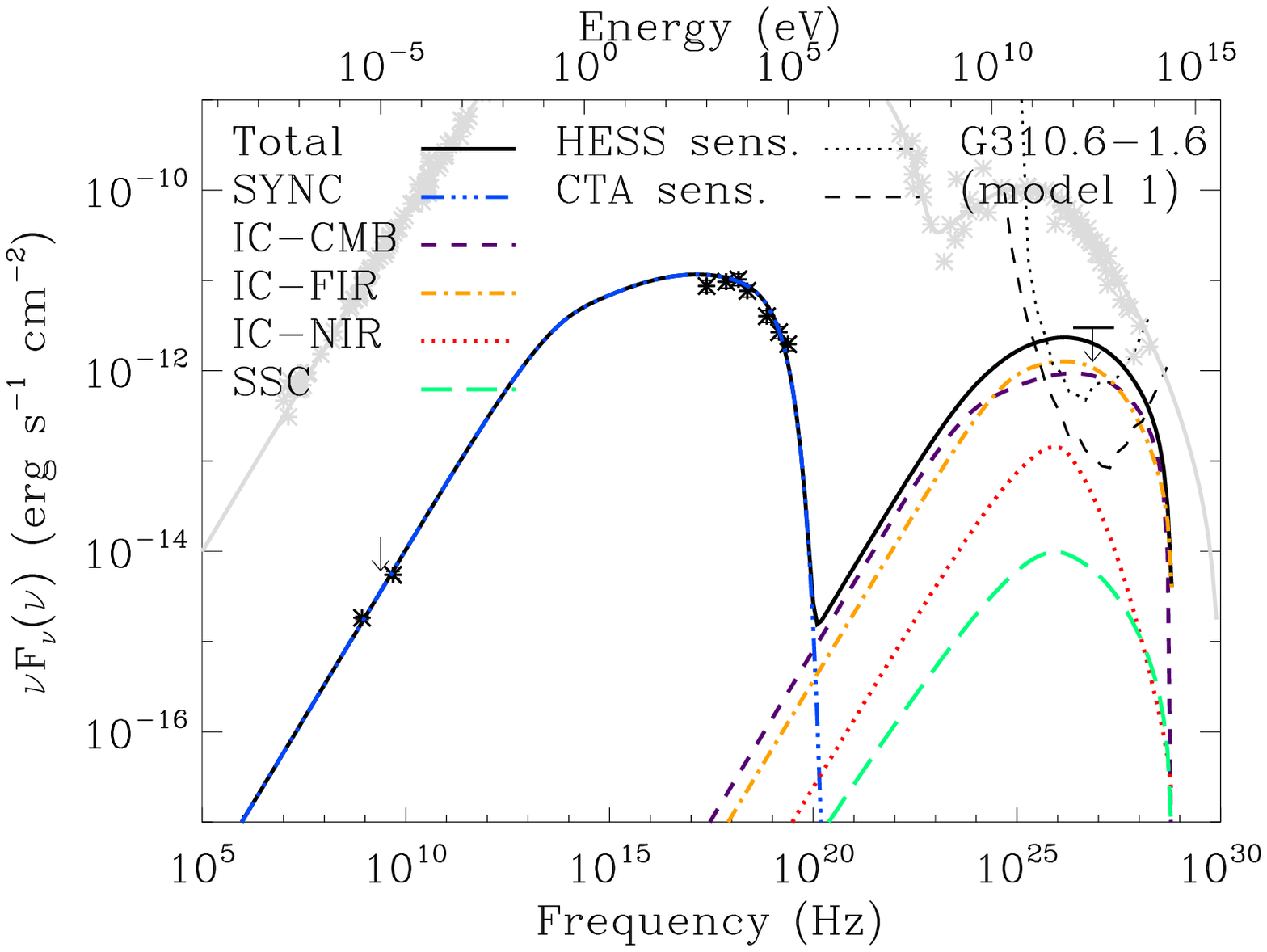}
%\hspace{0.8cm}
\includegraphics[scale=0.46]{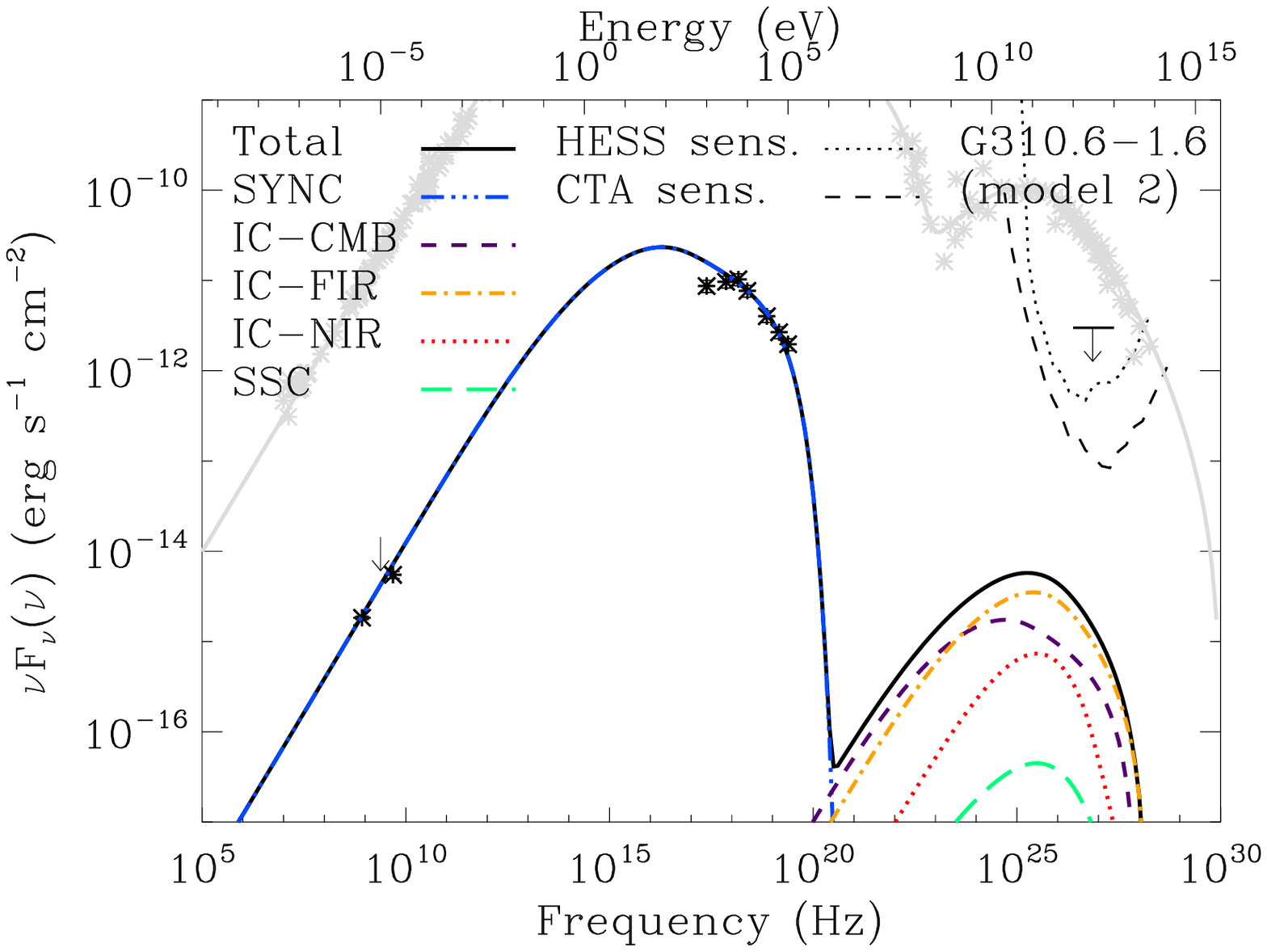}
\caption{Spectral fits for G310--1.6 PWN (models 1 to 2, top to bottom). }
\label{g310fig}
\end{center}
\end{figure}

The lack of observational constraints also allows considering an alternative model in which the nebula is a
magnetically dominated PWN with $\eta$=0.98, well beyond equipartition. In this case (model 2, see Fig. \ref{g310fig}),
the energy break moves to higher energies ($\gamma=6 \times 10^6$ or $\sim$3 TeV) and the magnetic field increases up to 306 $\mu$G.
Model 2 explains also well the overall X-ray flux, but fails in reproducing the break at 6 keV. 

Future observations of G310.6-1.6 will help to discern definitely between both models. For low-$eta$ model, the flux of G310.6-1.6 is a factor
$\sim$ over the H.E.S.S. sensitivity flux at 50 h of exposure time. Even with only the CMB contribution, this sensitivity is surpassed by a
factor $\sim$2. If the low-$\eta$ model is right, its detection is expected in a moderate exposure time with the current Cherenkov telescopes.

\subsection{G292.0+0.18}

As stated in the introduction, the pulsars related with G54.1+0.3 and G292.0+0.18 both have a period of  $\sim$135 ms, period
derivative of $\sim$ 7.5$\times10^{13}$, a spin-down power of 1.2$\times10^{37}$ erg/s, a characteristic age of $\sim$ 2900 years
and a distance of $\sim$6 kpc. For both pulsars, the braking index is unknown.

The radius of G292.0+0.18 is based on the SNR size of 8' diameter \citep{gaensler03}, which means a physical radius of 3.5 pc. The distance
estimate is based on the HI absorption profile given by \citet{winkler09}.
Based on measurement of the transverse motions of the filaments of the SNR and assuming that the shell is expanding with transverse
expansion velocity, \citet{winkler09} estimated an age between 3000 and 3400 years, concurring with \citet{gaensler03}. The ejected mass of the SN explosion was estimated as $\sim$6~M$_{\odot}$
\citep{gaensler03}.

Radio observations for the nebula were obtained from the work of \citet{gaensler03}. The flux of the nebula in X-rays was
measured by {\it Chandra} \citep{hughes01}. The photon index of the X-ray spectra, as it is suggested in \citet{hughes01}, is
considered the same as that of the pulsar. At GeV energies, we only have upper limits from Fermi-LAT \citep{ackermann11}. Optical
and near infrared observations were obtained for the torus of the nebula, by \citet{zharikov08} and \citet{zharikov13},
respectively, but these are not considered in our fits, since do not include the entire system. The background energy densities are
unknown. We assume those given by GALPROP, for which the equivalent temperatures and densities of the representing blackbodies are
T$_{FIR}$=25 K, $w_{FIR}$=0.42 eV cm$^{?3}$, and T$_{NIR}$=2800 K, $w_{NIR}$=0.70 eV cm$^{?3}$.

\begin{figure}
\vspace{0.2cm}
\begin{center}
\includegraphics[scale=0.46]{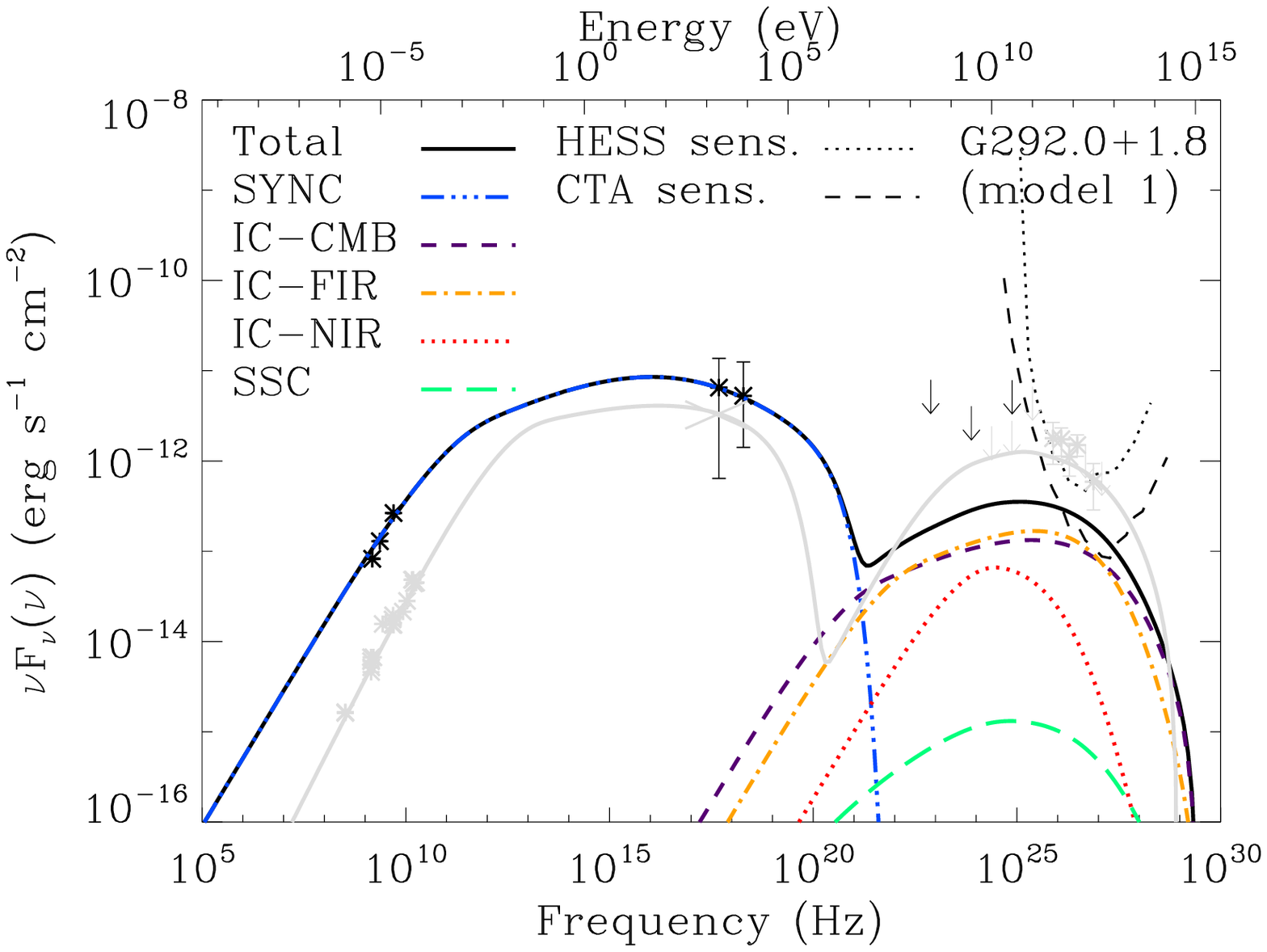}
\hspace{0.8cm}
\includegraphics[scale=0.46]{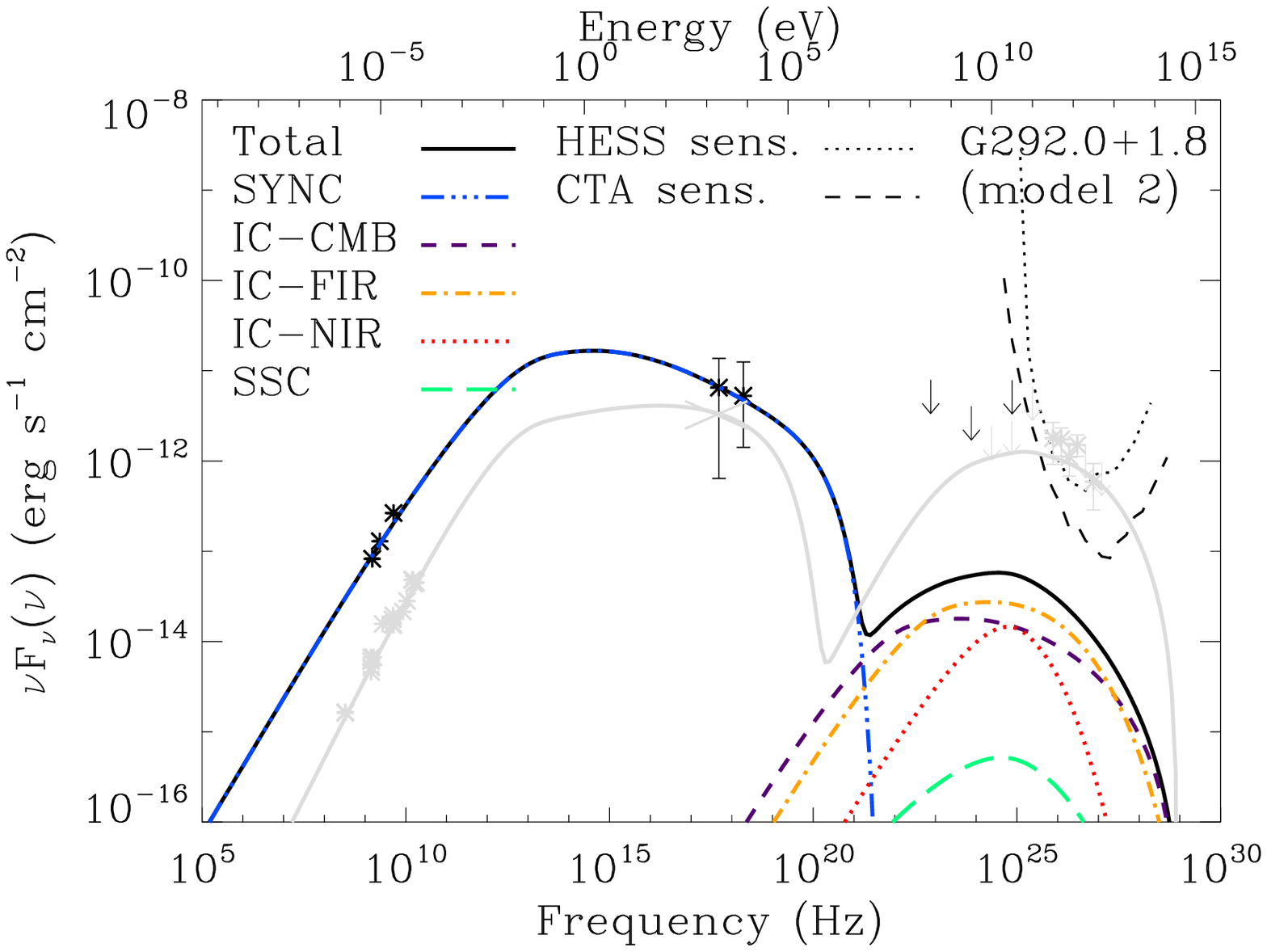}
\caption{Spectral fits for G292.0+0.18 PWN. In grey, we show the model and data for G54.1+0.3 extracted from \citet{torr14}.}
\label{g292fig}
\end{center}
\end{figure}

Fig. \ref{g292fig} shows two models that fit the radio and the X-ray data for this nebula. In both cases the age of the system
is 2500 years, and the ejected mass  is 9 M$_{\odot}$. In model 1 (see Fig.  \ref{g292fig}), we consider a low magnetic
fraction model with $\eta$=0.05, which is 10 times larger than the magnetic fraction obtained in our model for G54.1+0.3 in \citet{torr14}.
This model predicts that the nebula will be seen by CTA, and it would reach H.E.S.S. sensitivity if the FIR energy density reaches 2 eV
cm$^{-3}$. We obtain a magnetic field of 21 $\mu$G with an injection intrinsic break of $\gamma_b=10^5$ ($\sim$51 GeV), with a low (high)
energy index of 1.5 (2.55). These parameters differ from the ones obtained for G54.1+0.3 in \citet{torr14} ($B=14 \mu$G, $\gamma_b=5 \times
10^5$, $\alpha_l$=1.2, $\alpha_h$=2.8). With this model, the difference in the magnetic fraction, the energy densities of the IC target
photon fields and the age of the system explain why we observe G54.1+0.3 and not G292.0+1.8, even when both are particle dominated.

The radio and X-ray data are also compatible with a high-$\eta$ model for G292.0+1.8 (see Fig. \ref{g292fig})
with $\eta=0.77$ and a resulting magnetic field of 81 $\mu$G (similar to the Crab Nebula). The injection in this case has an energy break of
$\gamma_b=2.5 \times 10^5$ ($\sim$130 GeV) and the high energy spectral index changes slightly (2.5). In this case, G292.0+1.8
would not be detected even with CTA also explaining the difference with G54.1+0.3. A deep TeV observation will distinguish between these two
models. According to the first model (with $\eta=0.05$), the TeV flux would be only a factor $\sim$2 lower than the H.E.S.S. sensitivity limit in
50 h exposure time.

\section{Conclusions}

Despite having similar spin-down power,
the value of the magnetic
field differs from one to another PWN not only because of the value of $\eta$ 
($B\sim$ $\eta^{1/2}$) differs, but also because their size does
($B\sim R_{PWN}^{-3/2}$). 
Models with high values of $\eta$ would explain the low efficiency of some PWNe at X-rays and make
them undetectable at VHE. 
However, we here found that models with high magnetic field and fraction can be constructed only 
for some of the  nebulae that are
non-detected at TeV,
at the price of stretching other parameters. 
They
seem to work worse than particle dominated models in general, and remain viable only for G76.9+1.0 (for which
there are significantly less observational constraints) and G310.6--1.6 (pending the scrutiny of deeper TeV observations).
%
%From an strictly phenomenological viewpoint, the data now at hand in some cases can be explained either with a lot of particles and a very low magnetic field / magnetization, or with very little particles and a very high field / magnetization, but models are not equally satisfying. 
%
These are the specific conclusions:

\begin{itemize}

%\item We find two possible models for the TeV-detected N157B using data from radio, X-rays, and VHE and the estimations of age, radius and the energy
%density of possible target photon fields. Model 1  is compatible with the size of the nebula given
%by \citet{lazen00}, its age is compatible with the Sedov age of the remnant \citep{wang98}, the ejected mass is at the lower limit
%given by \citet{chen06}, and the energy densities of target backgrounds obtained are below the upper limits given by \citet{abra12}.
%This is a typical (luminous) TeV detected nebula, particle dominated. Model 2 fits the spectrum of N157B using the upper limits for the
%photon background densities given by \citet{abra12}, but it is necessary to assume a lower age (2500 yr). 
%Also here the nebula is particle dominated, and a high magnetization would render
%an impossible fit in both cases.

\item We propose a low magnetization model for N157B with an age of 4.6 kyr and a magnetic field of 13 $\mu$G. The size of the nebula
is compatible with the one given by \citet{lazen00}, the age with the Sedov age of the remnant \citep{wang98} and the ejected mass with the
lower limit given by \citet{chen06}. A high magnetization model ($\eta>0.5$) does not agree with the detection of N157B at TeV energies,
which would imply FIR and NIR energy densities much higher than the upper limits obtained by \citet{abra12}.

\item N158A non-detection seems to happen because of its smaller age
(perhaps also because of  a lower photon background?)
rather than by having a large magnetization. If this is the case, it will certainly be detected with CTA and likely also by the current
generation of instruments. Indeed, just the CMB inverse Compton contribution would produce a CTA source. Without taking into account a possible significant NIR contribution which would ease the required observation time,
we find that if N158A is subject to
a FIR energy density of 5 eV cm$^{-3}$, it can already lead to a detection by H.E.S.S. in 50 hours (lower IC target fields leads to larger
integration times, but still within plausible limits). The high-$\eta$
model(s) explored for N158A has been disregarded as unlikely due to inability to produce a good match to the X-ray data.

%\item In the cases of N157B and N158A, the SSC is not negligible, but not dominant. This non-dominance is
%consistent with the analysis of \citet{torr13}: Despite the great spin-down power of N157B the ages in model 1 and 2 are too large to
%find SSC dominance. In the case of N158A, where the assumed age is 760 yr, the spin-down power only represents 1/3 of the Crab nebula
%also making for a negligible SSC contribution. 

\item G76.9+1.0 is subject to a large uncertainty given the lack of sufficient observational constraints (only X-ray data are available).
This leads to the possibility of accommodating both extremes in the phase space of magnetic fraction. In none of the cases, a TeV detection is
expected and it will be difficult to differentiate among models. The FIR and NIR target fields necessary to reach the CTA sensitivity results in
more than a factor 100 (1000) in comparison with those obtained by GALPROP for model 1 (2). In such cases, the inverse Compton contribution at
X-ray energies would make impossible to fit the spectral slope. Other important parameters as age or the radius of the nebula are not well
determined and are needed to make a solid conclusion.

\item The low-$\eta$ models for G310.6--1.6 and G292.0+1.8 predict their detection with H.E.S.S. given sufficient integration time. The CMB
inverse Compton contribution reaches the sensitivity curve of a 50 hrs observation in the case of G310.6--1.6. The magnetic fraction for
G292.0+1.8 is one order of magnitude higher than the one obtained for G54.1+0.3 in \citet{torr14}. This fact and the slight difference in
the FIR and NIR energy densities considered in both cases, could explain the lack of detection of G292.0+1.8 at TeV. 
In both cases, radio
and X-ray data are also explained with a high-$\eta$ model with a magnetic field of 306 $\mu$G for G310.6--1.6 and 81 $\mu$G for G292.0+1.8.
However, the high-$\eta$ model for 
G310.6--1.6 is not preferred due to its inability to correctly reproduce the spectral break at 6 keV. For 
G292.0+1.8 instead, a high-$\eta$ model remains viable and TeV observations would solve the degeneracy.

\end{itemize}

\mbox{}\\
This work has been done in the framework of the grant AYA2012-39303. Furthermore, A.C.N. and D.F.T. acknowledge the grant PICT Ra\'ices
2012-0605.

\begin{table*}
\vspace{4cm}
\centering
\scriptsize
\vspace{0.2cm}
\label{tabfix}
\caption{Fixed or assumed model parameters.}
\begin{tabular}{lllllll}
\hline
Magnitude & Crab Nebula & N157B & N158A$^{\dag}$ & G76.9+1.0$^{\ddag}$ & G310.6--1.6 & G292.0+1.8\\
\hline
\hline
Pulsar magnitudes\\
\hline
$P$ (ms) & 33.40 & 16.12 & 50.50 & 48 & 31.18 & 135.48\\
$\dot{P}$ (s s$^{-1}$) & 4.21 $\times 10^{-13}$ & 5.18 $\times 10^{-14}$ & 4.79 $\times 10^{-13}$ & 8.64 $\times 10^{-14}$ & 3.89 $\times 10^{-14}$ & 7.53 $\times 10^{-13}$\\
$\tau_c$ (yr) & 1260 & 4936 & 1670 & 8970 & 12709 & 2854\\
$t_{age}$ (yr) & 940 & 4600 & 760 & 5000 & 1100 & 2500\\
$L(t_{age})$ (erg s$^{-1}$) & 4.5 $\times 10^{38}$ & 4.9 $\times 10^{38}$ & 1.5 $\times 10^{38}$ & 2.96 $\times 10^{37}$ & 5.1 $\times 10^{37}$ & 1.2 $\times 10^{37}$\\
$n$ & 2.51 & 3 & 2.08 & 3 & 3 & 3\\
$\tau_0$ (yr) & 730 & 336 & 2340 & 3970 & 11609 & 354\\
$d$ (kpc) & 2 & 48 & 49 & 10 & 7 & 6\\
$R_{PWN}$ (pc) & 2 & 10.6 & 0.7 & 4.7 & 1.3 & 3.5\\
\hline
Photon environment\\
\hline
$T_{FIR}^{(1)}$ (K) & 70 & 80 & 80 & 25 & 25 & 25\\
$T_{FIR}^{(2)}$ (K) & - & 88 & - & - & - & -\\
$T_{NIR}$ (K) & 5000 & - & - & 3200 & 3300 & 2800\\
\hline
Injection parameters\\
\hline
$\gamma_{min}$ & 1 & 1 & 1 & 1 & 1 & 1\\
\hline
\hline
\end{tabular}
\end{table*}

\begin{table*}
\vspace{4cm}
\centering
\scriptsize
\vspace{0.2cm}
\label{tabres}
\caption{Fitted or deduced model parameters.}
\begin{tabular}{lllllll}
\hline
Magnitude & Crab Nebula & N157B & N158A$^{\dag}$ & G76.9+1.0$^{\ddag}$ & G310.6--1.6 & G292.0+1.8\\
\hline
\hline
Pulsar magnitudes\\
\hline
$L_0$ (erg s$^{-1}$) & 3.1 $\times 10^{39}$ & 1.1 $\times 10^{41}$ & 3.3 $\times 10^{38}$ & 1.5 $\times 10^{38}$ & 6.1 $\times 10^{37}$ & 7.8 $\times 10^{38}$\\
$M_{ej}$ (M$_{\odot}$) & 9.5 & 20 & 25 & 20 & 9 & 9\\
\hline
Photon environment\\
\hline
$w_{FIR}^{(1)}$ (eV cm$^{-3}$) & 0.4 & 0.7 & 5 (0.2) & 0.13 & 0.62 & 0.42\\
$w_{FIR}^{(2)}$ (eV cm$^{-3}$) & - & 0.3 & - & - & - & -\\
$w_{NIR}$ (eV cm$^{-3}$) & 1 & - & - & 0.33 & 1.62 & 0.70\\
\hline
Injection parameters\\
\hline
$\gamma_{max}(t_{age})$ & 7.6 $\times 10^{9}$ & 3.8 $\times 10^{8}$ & 9.8 $\times 10^{8}$ & 5.7 $\times 10^{8}$ & 5.7 $\times 10^{8}$ & 2.4 $\times 10^{9}$\\
$\gamma_b$ & 7 $\times 10^{5}$ & $10^{6}$ & 3 $\times 10^{7}$ & $10^3$ & 2 $\times 10^{6}$ & $10^{5}$\\
$\alpha_l$ & 1.5 & 1.5 & 1.8 & 1.5 & 1.5 & 1.5\\
$\alpha_h$ & 2.5 & 2.75 & 2.6 & 2.7 & 2.5 & 2.55\\
$\epsilon$ & 0.25 & 0.02 & 0.3 & 0.25 & 0.3 & 0.3\\
\hline
Magnetic field\\
\hline
$B(t_{age}) ({\mu}G)$ & 82 & 13 & 32 & 3.5 & 8.2 & 21\\
$\eta$ & 0.02 & 0.006 & 0.0007 & 0.0017 & 0.0007 & 0.05\\
\hline
\hline
\multicolumn{7}{l}{
\begin{minipage}{14cm}
Some alternative models are commented in the text.
\end{minipage}
}\\
\multicolumn{7}{l}{
\begin{minipage}{14cm}
$^{\dag}$The FIR energy density in the table is the one  required for the PWN 
to be detected by H.E.S.S.  (CTA) in 50 hours.
\end{minipage}
}\\
\multicolumn{7}{l}{
\begin{minipage}{14cm}
$^{\ddag}$These parameters correspond to model 1 in Fig. \ref{g76}, other models are described in the text.
\end{minipage}
}
\end{tabular}
\end{table*}

\end{document}